\documentstyle [12pt]{article} 

\def\mcl{\mskip-8mu}
\def\defeq{\mathop{\stackrel{\rm def}{=}}\nolimits}  
\newtheorem{prop}{Proposition}

\begin{document}

\title{\vspace*{-\baselineskip}\bf Rapid solution of problems by
nuclear-magnetic resonance\\ quantum computation}   

\author{John M. Myers,$^*$\ \ A. F. Fahmy,$^{\dag}$\ \ S. J.
Glaser,$^{\ddag}$\ \ R. Marx$^{\S}$\protect\\
{\small $^*$Gordon McKay Laboratory,
Division of Engineering}\protect\\[-12pt] {\small and Applied Sciences,
Harvard University,  Cambridge, MA 02138, USA}\protect\\
{\small $^{\dag}$Biological
Chemistry and Molecular Pharmacology,}\protect\\[-12pt] {\small Harvard
Medical School, 240 Longwood Avenue, Boston, MA 02115, USA}\protect\\
{\small $^{\ddag}$Institut f\"ur Organische Chemie und
Biochemie, Technische}\protect\\[-12pt] {\small Universit\"at M\"unchen,
Lichtenbergstr.\ 4, D-85748 Garching, Germany}\protect\\
{\small $^{\S}$Institut f\"ur
Organische Chemie, J. W. Goethe-Universit\"at,}\protect\\[-12pt] {\small 
Marie-Curie-Str.\ 11, D-60439 Frankfurt, Germany}}

\maketitle 

\begin{abstract}We offer an improved method for using a
nuclear-magnetic-resonance quantum computer (NMRQC) to solve the
Deutsch-Jozsa problem.  Two known obstacles to the application of the
NMRQC are exponential diminishment of density-matrix elements with the
number of bits, threatening weak signal levels, and the high cost of
preparing a suitable starting state.  A third obstacle is a heretofore
unnoticed restriction on measurement operators available for use by an
NMRQC.  Variations on the function classes of the Deutsch-Jozsa
problem are introduced, both to extend the range of problems
advantageous for quantum computation and to escape all three obstacles
to use of an NMRQC.  By adapting it to one such function class, the
Deutsch-Jozsa problem is made solvable without exponential loss of
signal.  The method involves an extra work bit and a polynomially more
involved Oracle; it uses the thermal-equilibrium density matrix
systematically for an arbitrary number of spins, thereby avoiding both
the preparation of a pseudopure state and temporal averaging.
\end{abstract}

\newpage
\section*{\normalsize\bf INTRODUCTION}

Recently there has been interest in trying to use the thermal state as
a starting point for NMR computation.  We note two efforts to pursue
this, one by Zhou, Leung, and Chuang \cite{zhou}, the other by
Woodward and Br\"uschweiler \cite{woodward}.  We come at the problem
from a different point of view to obtain results slightly stronger
than those of \cite{zhou}, as well as showing some different ways to
proceed, and much more explicit than those claimed in \cite{woodward}.

Computational complexity brings the idea of cost {\em vs.}\/\ problem
size into problems solvable by use of computers.  For certain
problems, cost grows with problem size more slowly for quantum 
computers than it does for a Turing machine 
\cite{shor}\nocite{grover}--\cite{DJ}, showing that the complexity of a
problem depends on the computer used to solve it.  With the Turing machine
no longer the only game in town, the question is opened: what problems are
natural to one or another computer design \cite{adleman}? 

Are all quantum computers alike with respect to the problems that they
solve efficiently? Three types of quantum computer will be discussed
in connection with problems of function classification, the prototype
of which is the Deutsch-Jozsa (DJ) problem \cite{DJ}, which concerns
determining a property of an $n$-bit function $f:{\bf Z}_N \rightarrow
{\bf Z}_2$, given an oracle that evaluates $f$, where $N$ is written
as shorthand for $2^n$.

In theory, which is all this paper deals with, a quantum computer
yields the solution to a problem as the outcome of a quantum
measurement \cite{Deutsch1,Deutsch2}, and can be called an {\em
outcome} quantum computer (OQC) to distinguish it from an {\em
expectation-value} quantum computer (EVQC), which in place of an
outcome yields, to some finite precision, the expectation value for a
measurement operator and a (possibly mixed) state \cite{gradientPPS2}.
A {\em nuclear-magnetic-resonance quantum computer} (NMRQC) is a
restricted EVQC, the restriction stemming from facts of NMR
spectrometers.  The restriction on an NMRQC relative to a general EVQC
has consequences which seem to have gone unnoticed.  Attention to them
shows better how an NMR spectrometer can act as a quantum computer,
stimulates a generalization of the DJ problem, and shows the way to
solving the original DJ problem without exponential loss of signal as
the number of bits $n$ increases \cite{warren}.

\section*{\normalsize\bf DJ PROBLEM FOR THE OQC AND THE EVQC}

As stated originally, the DJ problem is this: given any function
$f:{\bf Z}_N\rightarrow {\bf Z}_2$, show at least one of following: 
(A) $f$ is not constant, or (B) $f$ is not balanced, 
where a balanced function has the value 0 for just half of its $N$
arguments and 1 for the other half.

We review briefly the history of methods for use of an OQC to solve
this problem.  For later generalization, it is convenient to organize
the method of solution in three steps, the middle one of which is a
compound step that may be repeated: (1) Prepare a quantum register in a
starting state; (2) apply operators including one for an Oracle for the
function $f$; and (3) make a quantum measurement defined by a
projection.

The method as first presented required a work bit and hence a quantum
register of $n+1$ bits; it also required two invocations of the oracle
(repetition of step (2)).  Later Cleve {\em et al.}\ showed how to solve
the problem invoking the Oracle only once \cite{cleve}; building
on this, Collins {\em et al.}\ showed how to skip the work bit so the
register is only $n$ bits \cite{collins}; this calls for a Hilbert space
spanned by $N$ orthonormal vectors $|j\rangle$, $j = 0, 1, \ldots, N-1$.
In this version, the method consists of the following steps:
\begin{enumerate}
\item Prepare the starting state \begin{equation}
|w\rangle \defeq N^{-1/2}\sum^{N-1}_{j=0}|j\rangle. \end{equation}  
\item Apply the operator $U_f$ for the Oracle for the function $f$
defined by its effect on basis vectors $|j\rangle$: \begin{equation}
U_f|j\rangle = (-1)^{f(j)}|j\rangle \label{eq:uf}\end{equation} (no
repetition and no other operators).
\item Make the measurement defined by the projection $|w\rangle\langle
w|$, which has eigenvalues 0 and 1, and hence two possible
outcomes. \end{enumerate} If the outcome is 1 the function is not
balanced, while if the outcome is 0 the function is not constant, as
follows from the probability of the outcome being 1: \begin{eqnarray}
\Pr\lbrace{\rm outcome} = 1\rbrace&\mcl=\mcl& {\rm Tr}(|w\rangle\langle
w|U_f|w\rangle\langle w|U_f^\dagger)\nonumber\\[4pt] 
 &\mcl=\mcl& \cases{1,&if $f$ is
constant,\cr 0,&if $f$ is balanced.\cr}  \end{eqnarray} (In
case $f$ is neither balanced nor constant, the OQC outcome can be
either 0 or 1 with probabilities determined by the usual rules of
quantum mechanics, but the outcome varies from one trial to another.)

Another version of the DJ problem restricts the class of functions to
be the union of constant and balanced functions; we shall have
occasion to introduce analogs to this version.

Turn now to the use of an EVQC, which in place of an outcome yields
the expectation value ${\rm Tr}(M \rho)$ for a measurement $M$ of a
density matrix $\rho$ \cite{gradientPPS2}.  An EVQC is characterized
by a parameter of resolution $\epsilon$: Two density matrices $\rho_1$
and $\rho_2$ are taken to be distinguishable by a measurement
described by an operator $M$ if and only if the difference in the
expectation values exceeds the minimum resolution: 

\begin{equation} |{\rm
Tr}(M \rho_1) - {\rm Tr}(M \rho_2)| > \epsilon
\Lambda(M),\label{eq:resolve}\end{equation} where $\Lambda(M)$ is the
difference between the minimum and the maximum eigenvalue of the
measurement operator $M$.  (The factor $\Lambda(M)$ makes limitations of
resolution immune to the mere analytic trick of multiplying the
measurement operator by a constant.) 
For an Oracle exercising $U_f$ on a density matrix $\rho$, a
measurement operator $M$ yields an expectation value \begin{equation}
E(f) = {\rm Tr}(M U_f \rho U_f^\dagger).\label{eq:tr}\end{equation}
Using the measurement operator $|w\rangle\langle w|$, an EVQC
measuring the state $U_f|w\rangle$ obtains the expectation value
\begin{eqnarray} E(f) &\mcl=\mcl& {\rm Tr}(|w\rangle\langle
w|U_f|w\rangle\langle w|U^\dagger_f) \nonumber \\ &\mcl=\mcl& N^{-2}
\sum_{j,k=0}^{N-1}(-1)^{f(j)+f(k)}.\label{eq:ef}\end{eqnarray} For
this case, it follows that \begin{equation} E(f) = \biggl( N^{-1}
\sum_{j=0}^{N-1} (-1)^{f(j)}\biggr)^2. \end{equation} This expectation
$E(f)$ has the nice property of invariance under permutations
of the arguments of $f$, and hence depends only on what might be
called the ``imbalance'' of $f$, defined by \begin{eqnarray} I(f) &\mcl
\defeq \mcl & \frac{1}{2}\,[\mbox{(Number of values
of}\ j \mbox{\ for which}\ f(j) = 1) \nonumber\\ && \mbox{} -(\mbox{Number
of values of}\ j \mbox{\ for which}\ f(j) = 0)].
\end{eqnarray}
Depending on $f$, $\,I(f)$ takes on integral values
$-N/2 \leq I(f) \leq N/2$.  (Recall $N=2^n$, and $n > 0$, so $N$ is
even.) That is, one has for this case \begin{equation} E(f) = 4
N^{-2}I^2(f).\label{eq:isq}\end{equation} For example, if $f$ is
balanced, one sees $I(f) = 0$, so it follows that $E(f)=0$, while if
$f$ is constant, $I(f) = \pm N/2$ so $E(f) = 1$; the two cases are
resolvable by an EVQC for any $\epsilon < 1$.

Drastic sensitivity to $\epsilon$ is seen in the satisfiability
problem of distinguishing the unsatisfiable function $f_0$ having the
zero value for all arguments from any function $f_1$ that takes the
value 1 for just one argument.  One can check to see that $I(f_0)=
-N/2$ and $I(f_1) = 1 - N/2$, so that
\begin{eqnarray}E(f_0)-E(f_1)&\mcl=\mcl&4 N^{-2}[(-N/2)^2 - 
(1-N/2)^2]\nonumber\\
&\mcl =\mcl& 2^{2-n}(1-2^{-n}). \end{eqnarray} This becomes
exponentially small as the number $n$ of bits increases, so that
$|E(f_0) - E(f_1)| > \epsilon\Lambda(|w\rangle\langle w|)$ only for
\begin{equation} n < \log_2(4/\epsilon).\end{equation}

\section*{\normalsize\bf GENERALIZATION}   

Equation (\ref{eq:ef}) suggests the following generalization.  Given any
$N\times N$ matrix $B$, define a mapping $S_B$ from the set of 
functions to numbers by \begin{eqnarray} S_B(f) &\mcl \defeq \mcl&
\sum_{j,k=0}^{N-1}(-1)^{f(j)+f(k)} B_{jk} \nonumber \\ &\mcl= &\mcl {\rm
Tr}(B) + \sum_{j=0}^{N-2} \sum_{k=j+1}^{N-1}(-1)^{f(j)+f(k)} (B_{jk} +
B_{kj}).
\label{eq:sym} \end{eqnarray} Then Eq.\
(\ref{eq:ef}) is equivalent to $E(f) = S_{B'}(f)$, where $B'$ is the matrix
defined by
\begin{equation}(\forall j,k)\ B'_{jk} = N^{-2} .\label{eq:sb1}
\end{equation} In the general case defined by (\ref{eq:tr}), one finds
\begin{eqnarray}E(f) &\mcl=\mcl& {\rm Tr}(M U_f \rho U_f^\dagger) \nonumber
\\ &\mcl=\mcl& \sum_{j,k=0}^{N-1}M_{jk}(-1)^{f(j)}\rho_{kj}(-1)^{f(k)}
\nonumber \\ &\mcl=\mcl& S_{B}(f) \label{eq:es} \end{eqnarray} for a matrix
$B(\rho,M)$ having elements \begin{equation} B_{jk} = M_{jk}\,
\rho_{kj} \;\;({\rm no \; sum}).\label{eq:sb}\end{equation} One is
thus led to explore generalizations of EVQC computations that
implement $S_B$ for matrices $B(\rho,M)$ of the general form of Eq.\
(\ref{eq:sb}) rather than the special form of Eq.\ (\ref{eq:sb1}).  In
particular, if $S_B(f)=0$, we shall say that $f$ is {\em balanced with
respect to}~$B$.  By inspection, one arrives at the following:
\begin{prop}For $\lbrace c_j\rbrace$ any set of constants and $\lbrace
B_j\rbrace$ any set of 
$N\times N$ matrices, if $f$ is balanced with respect to $B_1, B_2, \ldots,$
then $f$ is balanced with respect to $\sum_j c_j B_j$. \end{prop}

It follows from Eq.\ (\ref{eq:sym}) that \begin{prop}If the matrix $B$
is written as the sum of symmetric and antisymmetric parts, only the
symmetric part contributes to $S_B$.\end{prop}

For any $f:{\bf Z}_N \rightarrow {\bf Z}_2$, let $\bar{f}$ be the
logical complement of $f$, so $(\forall j)\ \bar{f}(j) = 1- f(j)$.
Then it follows immediately from Eq.\ (\ref{eq:uf}) that
\begin{prop}If $\bar{f}$ is the logical complement of $f$, then
\begin{equation} (\forall B)\ S_B(\bar{f}) = S_B(f).
\end{equation}
\label{prop:comp}\end{prop}\vspace{-24pt}

The three-step procedure for solving the DJ problem readily
generalizes to execute $S_B(f)$ for a variety of matrices $B$, as will
be illustrated in connection with the NMRQC.

\section*{\normalsize\bf NMR SPECTROMETER USED AS A COMPUTER}   

We review the use of a nuclear-magnetic-resonance spectrometer as an
NMRQC for solving the Deutsch-Jozsa problem, in order to point out
obstacles that impede it (relative to a general EVQC.  For step (1) on
an NMR spectrometer, a liquid sample begins in a mixed state of
thermal equilibrium and is manipulated one way or another into a
starting state.  The thermal-equilibrium density matrix is
proportional to $\exp(-{\cal H}/k_BT)$, where $\cal H$ is the
hamiltonian for the $n$-spin molecule (in the liquid sample) used as a
quantum register, $k_B$ is Boltzmann's constant, and $T$ is the
temperature.  In the high-temperature approximation the thermal
density matrix is given adequately well by the first two terms in the
Taylor expansion: \begin{equation} \rho_{\rm eq} = 2^{-n}({\bf 1} -
{\cal H}/k_BT) \approx N^{-1}{\bf 1} - \frac{\hbar}{N k_B T}\,\sum_i
\omega_i I^i_z,
\label{eq:therm}\end{equation} where $\omega_i$ is the resonant
angular frequency of the $i$-th nucleus, and $I^i_z$ is defined by a
tensor product over all $n$ spins in which all the factors are unit
operators except for $\frac{1}{2}\,{\rm Diag}(1,-1)$ as the $i$-th
factor of the tensor product.  This state, being diagonal, is
invariant under the action of the Oracle and so must be manipulated
into some other density matrix to serve as a starting state.

How to produce a starting density matrix has been much discussed.  One
way to prepare a starting density matrix is to produce a pseudopure
state using gradient pulses \cite{gradientPPS1,gradientPPS2},
resulting in a starting density matrix of the form \begin{equation}
\rho = (1- \alpha /N)N^{-1}{\bf 1} + \frac{\alpha}{N}\,|w \rangle
\langle w|, \label{eq:pps}\end{equation} for some (usually small)
coefficient $\alpha$; a cost is a reduction exponential in $n$ of
$\alpha$ and hence of the available spectrometer signal.  (The small
size of $\alpha$ compounds the exponential loss of polarization
expressed by the explicit appearance of $N$ in the formula for the
pseudopure state.)  Another way to deal with a starting state is
temporal averaging, which avoids the signal loss of a pseudopure
state, but requires repetitions of the whole procedure and addition of
the resulting spectra, costing much
time \cite{gradientPPS1}\nocite{temporalPPS}--\cite{marx}.  A third way
uses extra qubits as ancilla \cite{logicalPPS}, and a fourth advocates
another use of extra bits \cite{vazirani}.  All these methods are elaborate
and expensive of signal or time or number of bits required.  A ray of hope
is the simplified use of the equilibrium density matrix, which has been
shown to work for the DJ problem for functions of one bit \cite{chuang2} and
two bits 
\cite{linden}, but has not been developed into an algorithm applicable to
the general case of $n$ bits.

Whatever method prepares a starting state, in step (2) a unitary
transformation on the density matrix is implemented by use of
r.f.\ pulses combined with waiting periods during which spin-spin
couplings inherent in the molecule of the liquid sample exercise their
effect.  This results in some density matrix $\rho'$ at some time
$t'$.

Step (3), which we particularly want to notice, is modified in NMR to
result in a spectrum conventionally expressed as the time evolution of
the measurement of $F^+$, which is equivalent to the simultaneous
measurement of $F_x$ and $F_y$, defined by \begin{equation} F_{x,y} =
\sum_{j=1}^n I^j_{x,y},\end{equation} where $I^j_x$ is a tensor
product over all $n$ spins in which all the factors are unit operators
except the $j$-th factor, which is \begin{equation} I_x = \frac{1}{2}
\left( \matrix{0&1\cr 1&0\cr} \right); \end{equation} $I^j_y$ has
instead of $I_x$ the $j$-th factor \begin{equation}I_y = \frac{1}{2}
\left( \matrix{0&-i\cr i&0\cr} \right). \end{equation} 
(If the resonances of individual spins $j$ are well resolved (e.g. if
spin $j$ has a unique gyromagnetic ratio),the corresponding
$I_{x,y}^j$ can be measured using analogue or digital filters, and not
just the sum over all $j$.)  The spectrometer signal for $F_x$ starts
at $t'$ and is a sequence of expectation values obtained at
measurement times $t_k = t' + k \Delta t$, $k = 0, 1, 2$, \dots, where
$\Delta t$ is the sampling interval.  In the Heisenberg picture, the
density matrix $\rho'$ is fixed and the $k$-th expectation value is
${\rm Tr}(\rho' M_k)$, where, for example with $M = F_x$,
\begin{equation} M_k = \exp\biggl(\frac{i}{\hbar}\,k\Delta t 
{\cal H}\biggr) F_x \exp\biggl(-\frac{i}{\hbar}\,k \Delta t{\cal
H}\biggr).\label{eq:data}\end{equation} Analogous time sequences can
be defined for $F_y$ and, in the well resolved case, for $I^j_{x,y}$.

Typically, the signal (which is damped by relaxation in a way not  
shown in Eq.\ (\ref{eq:data})) is Fourier transformed into an NMR
spectrum. Either one deals with complications from a less than general
coupling, e.g., by use of swap operations \cite{Brueschweiler}, or one
must use a molecule and a spectrometer which exhibit distinct
frequencies for all single-spin transitions.

\vskip7pt
\noindent {\bf Remark}: For a 
molecule in which all transitions between basis states have distinct
frequencies, to see them one must resolve all $n 2^{n-1}$ peaks of the
Fourier spectrum, which requires a time-bandwidth product exponential
in the number of spins.

\vskip7pt
\noindent This makes it desirable to avoid Fourier
transforms of the time-domain signal, leading us to focus on single-time
measurements which involve no Fourier transform.  The requirement that
single-time measurement operators in NMR be unitarily equivalent to
$F_{x,y}$ or to $I_{x,y}^j$ now becomes an obstacle, because the
operators $F_{x,y}$ and $I_{x,y}^j$ all have spectra with multiple
eigenvalues, so that no single-time operator is nondegenerate.  Thus
no single-time operator has the power of a nondegenerate operator to
resolve states; this constraint limits the NMRQC.

The original method for solving the DJ problem used for its
measurement the projection operator $|w,0\rangle\langle w,0|$ while
the streamlined method used $|w\rangle\langle w|$, also a projection
operator.  Less important than it seems at first glance but still
provoking of thought is the following: \begin{prop}No single-time
measurement of any NMR operator can implement any nontrivial
projection.\end{prop} {\bf Proof}: Any single-time operator $M_k$ is
some unitary transform of some weighted sum of operators $I^j_{x,y}$,
all of which are traceless.  Trace is preserved under unitary
transform, so all the candidates for $M_k$ are traceless (which indeed
they must be if the large term proportional to the unit matrix in
Eq.\ \ref{eq:therm} is not to drown out all the effects of interest). A
nontrivial projection has nonzero trace.  Q.E.D.

\vskip7pt
\noindent One can get
around Proposition 4 by invoking a nonprojective operator to
distinguish balanced from constant functions, but questions remain
that are less easily disposed of.  One requires in place of
$|w\rangle\langle w|$ an operator $M$ that (a) works with the starting
density operator of the form of Eq.\ (\ref{eq:pps}), and (b) via
Eq.\ (\ref{eq:ef}) produces an expectation value that is invariant under
permutation of the arguments of $f$.  It is proved in 
Appendix A that:

\begin{prop} (i) Given a density operator $\rho$ of the form of Eq.\
(\ref{eq:pps}), the expectation value ${\rm Tr}(M U_f \rho U_f^\dagger)$ is
invariant under permutation of the arguments of $f$ if and only if $M =
c|w\rangle \langle w| + D + A$, where $D$ is any diagonal matrix, $\,A$ is
any antisymmetric matrix, and $c$ is any scalar; (ii) the resulting
expectation value $E(f)$ is independent of the antisymmetric matrix A;
(iii) if $M$ is hermitian, $c$ and $D$ are real and $A$ is pure imaginary. 
\label{prop:form}\end{prop}

Using a measurement operator $c |w\rangle \langle w| + D + A$
unitarily equivalent to $F_x$ and the starting density matrix of Eq.\
(\ref{eq:pps}), the expectation value for constant functions differs
from that for balanced functions by $c\alpha/N$, with the result that
the two classes of functions are distinguishable if and only if the
resolution satisfies
\begin{equation} \epsilon < \frac{\alpha}{N}\,|c|/\Lambda(F_x) =
\frac{\alpha |c|}{n N} ,\label{eq:cn}\end{equation}
where the second equality follows from $\Lambda(F_x) = n$ as the
difference between the minimum and maximum eigenvalues of $F_x$.
Thus a small value of $|c|$ demands fine resolution.

The straightforward way to produce a measurement operator in NMR
spectrometry is by unitary transform of $F_x$.  (It adds nothing to
allow unitary transforms of $F_y$, which is unitarily equivalent to
$F_x$.)  This and Eq.\ (\ref{eq:cn}) raise the question of how large a
value of $|c|$ is possible for an operator of the form $M = c
|w\rangle \langle w| + D + A$ that is constrained to be unitarily
equivalent to $F_x$.  As follows from the invariance of eigenvalues
under unitary transform, the constraint is that $F_x$ and $M$ have the
same eigenvalues with the same multiplicities.  This implies
\begin{prop} For the matrix $M$ of Proposition \ref{prop:form} to be
unitarily equivalent to $F_x$, it is necessary that $D$, $\,A$, and $c$ be
such that for all eigenvalues $\lambda_k$ of $F_x$, $\,k = 1$, \dots,\ $N$,
$\,\det(M-\lambda_k) = 0$ and ${\rm Tr}(M) = {\rm Tr}(F_x) = 0$.
\end{prop}
It is instructive to 
look at the first two cases, $n=1$ and $n=2$. For $n=1$, one has $F_x
= I_x$ and $M = I_x$ produces the largest possible value of
$|c|/\Lambda(M)$ consistent with the eigenvalues of $\pm 1/2$, namely,
$\,|c|/\Lambda(M) = 1$.  For the two-spin case ($n=2$), an analysis of the
restriction that the eigenvalues be those of $F_x$ shows
\begin{equation} {\rm For \;} n =2,\; |c|/\Lambda(M) < (2/3)^{1/2};
\end{equation} with more work, somewhat lower bounds can be
demonstrated.  (We found an $M$ for two spins unitarily equivalent to
$F_x$ for which $|c|/\Lambda(M) = 3^{-1/2}$, but we do not know if
this is the best that can be done.)  Thus there is a drop-off in
$|c|/\Lambda$ between the case $n=1$ and the case $n=2$, and hence an
increase in the fineness of required resolution relative to $\alpha/N$
(see Eq.\ (\ref{eq:cn})).  This drop-off suggests the following
question for future analysis:

\vskip7pt
\noindent {\bf Question}: For a measurement
operator of the form $M = c |w\rangle \langle w| + D + A$, unitarily
equivalent to $F_x$, how does the largest possible value of
$|c|/\Lambda(M)$ vary with the number of bits $n$?

\vskip7pt
\noindent Via Eq.\ (\ref{eq:cn}), the answer to this question will
determine as a function of $n$ the resolution necessary for an NMRQC to
solve the DJ problem using this method.  The cases examined suggest a
decreasing function; if confirmed this poses a serious obstacle of signal
loss beyond that already known to the use of this method to solve the DJ
problem.

\section*{\normalsize\bf EXAMPLES OF FUNCTIONS NATURAL TO NMR}  

With this background, we ask: are there function classes for which a
single-time measurement suffices to distinguish a function of that
class from the constant function? Here are some such classes, the
definitions of which depend on the concept of a Hamming distance.  Let the
argument $j$ of a function
$f:{\bf Z}_N \rightarrow {\bf Z}_2$ be written as an $n$-bit string,
padded with 0's to the left.  Given two integers $j,k$ (with $0 \leq
j,k \leq 2^n-1$), let $d$ be the number of bits of $j$ that are
different from the corresponding bits of $k$.  This is the Hamming
distance between $j$ and $k$, denoted $d(j,k)$.  Consider functions
$f:{\bf Z}_N \rightarrow {\bf Z}_2$ such that: a) $f(j) = 1$ for $N/4$
values of $j$, and b) if $f(j) = f(k) = 1$, then $d(j,k) \neq 1$.  Let
${\cal C}_N$ be the set of all such functions together with all their
binary complements.

\begin{prop} For all $j$, every function of ${\cal C}_N$ is balanced
with respect to $I_x^j$.\end{prop} {\bf Proof}: Suppose $f \in {\cal
C}_N$; let $g$ be $f$ or $\bar{f}$, whichever function takes the value 1
for $N/4$ of its arguments.  By Proposition \ref{prop:comp} it
suffices to show that, for all $j$, $U_g I^j_x U^\dagger_g = 0$.  The
($l,m$)-element of $I_x^j$ is nonzero if and only if the ($n$-bit
representations of) $l$ and $m$ differ at just bit $j$.  Hence this
element is nonzero only if the Hamming distance $d(l,m)=1$.  Because
$I_x^j$ is proportional to a permutation matrix, it has one nonzero
element in each row, so $U_g$ acting on its left changes the sign of
half the nonzero elements of $I_x^j$.  $U_g$ multiplied on the right
also changes half the nonzero elements of $U_g I_x^j$.  If elements
negated by multiplication on the right are distinct from those negated
by multiplication on the operation on the left, then half the elements
change sign and we are done.  For this to fail, at the element
$(l,m)$, it must be that $g(l)=g(m)=1$ and $(I_x^j)_{l,m} \neq 0$.
But that can happen only for functions not in ${\cal C}_N$.  Q.E.D.

From this proposition, it will be shown that functions of ${\cal C}_N$
can be efficiently distinguished from constant functions by use of an
NMRQC; moreover, solving this problem by use of a classical computer
requires a number of function evaluations that grows exponentially
with $n$.

Consider a starting state $\rho$ prepared from the equilibrium 
density operator by a hard $90^\circ$ $y$-pulse: \begin{equation}
\rho \defeq U_{90y}\rho_{\rm eq}U^\dagger_{90y} \approx
N^{-1}{\bf 1} - \frac{\hbar}{N k_B T}\,\sum_{i=1}^n \omega_i I_x^i
;\label{eq:start}\end{equation} suppose the Oracle executes $U_f$ and
the measurement operator is $F_x = \sum_i I_x^i$.  Applying Eq.\
(\ref{eq:sb}) to this case and using $(I_x^j)_{l,m}(I_x^k)_{m,l} =
\delta_{jk}/4$, one finds \begin{equation} B(\rho,F_x) =
- \frac{\hbar}{2 N k_B T}\,\sum_i \omega_i I^i_x.\end{equation}

By Propositions 1 and 7, every function $f \in {\cal C}_N$ is balanced
with respect to $B$, so $S_B(f) \equiv E(f) = 0$, regardless of
$\omega_i$.  In contrast, for the constant functions $f_0(j) = 0$ and
$f_1(j) = 1$ (for all $j$), one finds \begin{equation} E(f_{0,1}) =
\sum_{l,m=0}^{N-1} B_{lm} = \frac{\hbar}{k_B T}\,\sum_{i=1}^n \omega_i .
\end{equation}  Notice the absence of a factor of $N$ in the
denominator, removed by summing over the $N$ elements of $I_x^i$, each
1/2.  Hence, neglecting effects beyond reach of this theory, an NMRQC
operating with the starting density matrix defined in Eq.\
(\ref{eq:start}) can distinguish, for any $n$, functions of class
${\cal C}_N$ from constant functions for any resolution
\begin{equation} \epsilon < \frac{\hbar}{n k_B T}\,\sum_{i=1}^n \omega_i
.\end{equation} A striking feature of this result is the appearance in
the denominator of $n$, the number of nuclear spins, rather than
$N \equiv 2^n$.  Hence we have a method that avoids the much lamented
exponential loss of signal.

\section*{\normalsize\bf THE THERMAL STATE AND $|w\rangle\langle
w|$-BALANCED FUNCTIONS}

At the expense of an extra bit and a more complex Oracle, the balanced
functions (i.e., balanced with respect to $|w\rangle\langle w|$) can be
distinguished from constant functions using the starting state
obtained merely by a hard $90^\circ$ $y$-pulse applied to the thermal
state (see Eq.\ (\ref{eq:start})), requiring neither the pseudopure
state of Eq.\ (\ref{eq:pps}) nor temporal averaging.  One requires an
Oracle for a function $f:{\bf Z}_{N/2} \rightarrow {\bf Z}_2$ that
implements $U_{f'}$ not for $f$ but for $f':{\bf Z}_N \rightarrow {\bf
Z}_2$, related to $f$ by \begin{equation} f'(j) = \cases{f(j),&if $0\leq 
j \leq N/2 - 1$,\cr 0,&if $N/2 \leq j \leq N-1$.\cr}
\end{equation} Thus while the function $f$ is a function on $n-1$ bits 
balanced with respect to $|w\rangle \langle w|$, $f'$ is a function on
$n$ bits balanced with respect to $I^1_x$.  The three steps of
execution to decide if $f$ is balanced or constant are then: (1) apply
a hard 90$^\circ$ $y$-pulse on the thermal state, (2) apply $U_{f'}$
for $f'$ related to $f$ as above, (3) measure $I^1_x$ in the time
domain, in the limit of small times, to obtain a signal that is
substantial if $f$ is constant but vanishes if $f$ is balanced with
respect to $|w\rangle \langle w|$.

It is easy to check that given this
more complex Oracle, an NMRQC decides between balanced functions and
constant functions for any resolution
\begin{equation} \epsilon < \frac{\hbar}{k_BT}\,\omega_1
.\end{equation} There is no exponential growth in the demand for
resolution; indeed there is no growth at all, an advantage over the
procedure described in \cite{zhou}.  (The factor of $n$ in Ex.\
(\ref{eq:cn}) vanishes when $F_x$ is replaced by $I_x^1$.) This shows
a way to solve the original DJ problem on NMR with no loss of signal
as $n$ increases.

It should be remarked that the operations discussed do not display the
couplings needed to make any kind of quantum computer serve to
distinguish the function classes discussed.  These couplings are
required, however, in the NMR implementation of the Oracle.

\appendix
\section*{\normalsize\bf APPENDIX A.  PROOF OF PROPOSITION \ref{prop:form}} 

Assume for some $\alpha$ that $\rho = N^{-1}(1-\frac{\alpha}{N}){\bf
1} + \frac{\alpha}{N}\,|w\rangle \langle w|$. Then we have for the
expectation value, $E(f) = {\rm Tr}(M \rho) = 
N^{-1}[(1-\frac{\alpha}{N})\,{\rm Tr}(M) + \frac{\alpha}{N}\,S_M(f)]$,
whence it follows that $(\forall M, f,g)\ [E(f) = E(g)
\Longleftrightarrow S_M(f) = S_M(g)]$.  To see the condition imposed
on $M$ by the required invariance of $S_M$ under permutations, let
$P_{lm}$ be the matrix obtained by permuting rows $l$ and $m$ of the
$N\times N$ identity matrix.  Define an operation of $P_{lm}$ on $f$ by
\begin{equation} (P_{lm}f)(j) = \cases{f(m),&
if $j = l$,\cr f(l),&if $j = m$,\cr f(j),&otherwise.\cr}
\label{eq:pf} \end{equation}

Because general permutations are compositions of elementary
permutations, the necessary and sufficient condition for ${\rm Tr}(M\rho)$
to be invariant under all permutations is that
\begin{equation}(\forall l,m, f)\ S_M(f) =
S_M(P_{l,m}f).\label{eq:slm}\end{equation} It follows from Eqs.\
(\ref{eq:sym}) and (\ref{eq:pf}) that for any $f$ such that $f(l) \neq
f(m)$, 
\begin{eqnarray} \lefteqn{S_M(P_{lm}f)}\nonumber\\ &\mcl =\mcl& {\rm Tr}(M)
+\sum_{j=0}^{N-2}
\sum_{k=j+1}^{N-1}(-1)^{P_{lm}f(j)+P_{lm}f(k)}(M_{jk} + M_{kj})
\nonumber \\ &\mcl=\mcl& {\rm Tr}(M) + \sum_{j=0}^{N-2}
\sum_{k=j+1}^{N-1}(-1)^{f(j)+f(k) + \delta{jl} + \delta_{jm}+
\delta_{kl} + \delta_{km}}\nonumber\\
&&\mbox{}\times(M_{jk} + M_{kj}) ,
\label{eq:smp} \end{eqnarray} where $\delta_{jl}$ is the Kronecker
$\delta$, equal to 1 if $j=l$ and otherwise equal to 0.

For any $l < m$, assume any $f$ such that $f(l) \neq f(m)$; set
$\hat{M}_{jk} = M_{jk} + M_{kj}$, require $S_M(f) = S_M(P_{lm}f)$, use
Eq.\ (\ref{eq:smp}), and eliminate terms that are the same on the two
sides of the equation to show: \begin{eqnarray}0 &\mcl=\mcl&
\sum_{k=m+1}^{N-1}(-1)^{f(k)}(\hat{M}_{lk} - \hat{M}_{mk})\nonumber \\
& &\mbox{} +
\sum_{j=0}^{l-1}(-1)^{f(j)}(\hat{M}_{jl} - \hat{M}_{jm}) \nonumber \\
& &\mbox{} + \sum_{k=l+1}^{m-1}(-1)^{f(k)}\hat{M}_{lk} -
\sum_{j=l+1}^{m-1}(-1)^{f(j)}\hat{M}_{jm}. \label{eq:symp}
\end{eqnarray} (The convention is used that if the upper limit of a
sum is less than the lower limit, the sum is zero.) On relabeling some
indices in the sums, this becomes \begin{eqnarray} \lefteqn{(\forall l <
m)(\forall f\ {\rm s.t.}\ f(l) \neq f(m))}\qquad\quad\nonumber\\
 0&\mcl=\mcl&
\sum_{k=0,k\neq l, k
\neq m}^{N-1} (-1)^{f(k)}(\hat{M}_{lk} - \hat{M}_{mk}). \end{eqnarray} This
can hold for all admissible $f$ only if \begin{equation} (\forall l <
m)(\forall k \neq l,m)\ \hat{M}_{lk} = \hat{M}_{mk}.\end{equation}
This, together with the symmetry from its definition that
$\hat{M}_{jk} = \hat{M}_{kj}$, implies that for $(\forall j \neq
k)\ \hat{M}_{jk}$ is independent of $j$ and $k$.  From this, part (i)
of the proposition follows immediately.  Part (ii) is an immediate
consequence of Proposition 2, and part (iii) depends only
on the definition of a hermitian matrix.  Q.E.D.

\section*{\normalsize\bf APPENDIX B.  DECIDING BETWEEN ${\cal C}_N$ AND
CONSTANT\protect\\ FUNCTIONS CLASSICALLY}  

\begin{prop} The number of invocations of a classical Oracle required
to decide with certainty that a function $f \in {\cal C}_N$ is not
constant is at least $2^{n-1}+1.$\end{prop} {\bf Remark}: The issue in
proving this is to rule out the possibility that the constraint on the
Hamming distance associated with the function class ${\cal C}_N$ can
greatly reduce the number of invocations required of the Oracle.

\vskip7pt
\noindent {\bf Proof}: Given an Oracle that, on demand, takes an argument
$j$ and computes the function value $f(j)$, how many invocations of the
Oracle are sufficient to assure a decision between ``$f$ is constant''
and ``$f \in {\cal C}_N$''?  Suppose one has obtained from the Oracle the
values $f(j)$ for any $K$ values of $j$, with $K \leq N/2$, and
suppose for all these arguments, $f(j) = 0$.  Then the possibility that
$f$ is constant is not excluded.  What about the possibility that $f
\in {\cal C}_N$?  We show that under these conditions there exists an $f'
\in {\cal C}_N$ that satisfies $f'(j) = 0$ for all the $K$ arguments
tested, so the possibility that $f \in {\cal C}_N$ is also not excluded.
This follows as soon as we show that in any set of $N/2$ arguments
there exist a subset of $N/4$ arguments each separated by a Hamming
distance greater than 1 from all the others of the subset.  To see
this is so, observe that out of the at least $N/2$ arguments unchecked
by the Oracle, any one can be chosen and called $j_0$.  Partition all
the arguments of the unchecked subset into classes $W_m$ where $k \in
W_m$ if and only if $m = d(j_0,k)$. Observe that $( \forall j,k)\ d(j,k)=1
\Rightarrow [( \exists m = m' \pm 1)$ such that $j \in W_m$ and $k \in
W_{m'}]$.  From this it follows that for any pair $j$ and $k$ both in
$W_0 \bigcup W_2 \bigcup$, \dots, $d(j,k) \neq 1$; similarly for any
pair $j,k$ both in $W_1 \bigcup W_3 \bigcup$, \dots, $d(j,k) \neq 1$.
At least one of these unions of $W$-classes has $N/4$ elements, and
hence holds arguments for some $f' \in {\cal C}_N$. Q.E.D.

\end{document}